\documentstyle[preprint,version2,aps]{revtex}

\newcommand{\Nu}[1]{\mbox{$n^\uparrow_{\vec{#1}}$}}
\newcommand{\Nd}[1]{\mbox{$n^\downarrow_{\vec{#1}}$}}
\newcommand{\Nt}[1]
{\mbox{$n_{\uparrow\downarrow}(t,\vec{r},\vec{#1})$}}

\begin{document}
\draft
\begin{quote}
\raggedleft cond-mat/9406101
\\RU-94-49
\end{quote}
\begin{title}
Low-Temperature Spin Diffusion in a
Spin-Polarized Fermi Gas
\end{title}

\author{D. I. Golosov and A. E. Ruckenstein}
\begin{instit}
Department of Physics, Rutgers University, Piscataway, NJ 08855-0849,
U.S.A.
\end{instit}
\begin{abstract}

We present a finite temperature
calculation of the transverse spin-diffusion coefficient,
$D_\bot$,
in a dilute degenerate
Fermi gas in the presence of a small external magnetic field, $H$.
While the longitudinal diffusion coefficient displays the conventional
low-temperature Fermi-liquid behavior, $D_\parallel \propto T^{-2}$,
the corresponding results for $D_\bot$ show
three separate regimes:
(a) $D_\bot \sim H^{-2}$ for $T \ll H$;
(b) $D_\bot \sim T^{-2}$, $D_\bot /D_\parallel \neq 1$ for $T \gg H$
and large
spin-rotation parameter $\xi \gg 1$, and (c) $D_\bot =
D_\parallel \propto T^{-2}$ for $T \gg H$ and $\xi \ll 1$.
Our results are qualitatively consistent with the available
experimental data in weakly spin-polarized
$^3{\rm  He}$ and $^3{\rm He} - ^4{\rm He}$ mixtures.
\end{abstract}

The unusual features of spin dynamics in spin-polarized quantum
systems have been intensively studied since the pioneering
paper of Leggett and Rice~\cite{Leggett} on spin
diffusion in normal liquid $^3{\rm He}$.
The main effect arises from the observation that the presence of a
molecular field (induced by the applied magnetic field) leads to
the an additional precession of the spin-current which in
steady state acquires a component perpendicular to the
magnetization gradient; through the continuity equation this results
in an anomalous reactive component (damped spin-wave) to spin transport.
This ``spin-rotation" effect is also present in the case of
spin-polarized Boltzmann gases~\cite{gases}.

{}From a microscopic point of view,
it was expected that a number of
qualitatively novel phenomena might arise in the case of
sufficiently high polarizations~\cite{Metamagnetism}.
A natural suggestion, made by Meyerovich~\cite{Meyer85}, was that
low-temperature spin-diffusion becomes highly anisotropic for finite
polarizations.
More precisely,
processes involving
spin-flips make use of the phase-space volume between the
two distinct
Fermi surfaces (for ``up" and ``down" spins) leading to
a {\em finite} scattering rate in the limit of $T\rightarrow 0$. This is
in contrast with processes involving scattering in the vicinity
of each of the Fermi surfaces which are subject to the
phase-space restrictions of unpolarized Fermi liquids and are thus
characterized by the
conventional Fermi-liquid behavior of scattering rates,
$\propto T^2$~\cite{Abrikosov}.
Meyerovich's suggestion was recently supported
by measurements of
the transverse spin-diffusion coefficient in weakly polarized liquid
$^3{\rm He}$~\cite{Candela}.
Theoretically, $D_\bot$ was calculated in the dilute gas limit
at $T=0$~\cite{Mullin89,KarenPRL} and
only a tentative estimate based on a variational solution of
the Boltzmann equation is available for $T\neq 0$ ~\cite{Mullin92}.

In this letter we present the analytical calculation of the finite
temperature behavior of $D_\bot$ for a dilute, weakly polarized Fermi gas,
in the s-wave
approximation by solving the appropriate kinetic equations exactly.
To the extent to which in the limit of small polarizations
strong interactions only lead to
constant renormalizations
of the weakly interacting result we expect that our findings
should
also apply to the case of weakly polarized $^3{\rm He}$. Indeed, as explained
below, the
detailed low $(H,T)$ behavior of our expression for $D_\bot$ is
consistent with the small systematic deviations (lying within the error-bars)
in the results of Ref. \cite{Candela}
from previous theoretical
estimates.
We note that the exact solution of the kinetic equations
does not merely lead to quantitative renormalizations of the transport
coefficients; rather, it brings out
new qualitative effects.
In particular, due to the existence of two dimensionless parameters,
the low-temperature behavior of transverse diffusion displays two
crossovers, the first at $T\sim H$ and the other at
$T \propto \sqrt {H\epsilon _F}$ ($\epsilon _F$ is the Fermi energy).
Our finding, not
appreciated in previous work on the subject,
is that spin diffusion becomes isotropic
(i.e., $D_\bot =D_{\parallel}$) only for $T$ above the {\em second}
crossover.

Our starting point is the kinetic equation for the Wigner transform
of the transverse component  of
the density matrix (in the frame rotating at the bare Larmor
frequency),
$n_{\uparrow \downarrow} (t, \vec{r}, \vec{p})$, where $t, \vec{r}$ are
the center of mass time and space coordinates and $\vec{p}$ is the
relative momentum vector:
\begin{eqnarray}
&&\left(\frac{\partial}{\partial t}
+\frac{\vec{p}}{m}\frac{\partial}{\partial \vec{r}}\right) \Nt{p} =
\Nt{p}\,  \int \left\{ i \left[\frac{4 \pi a}{m} +
B(\vec{p},\vec{p}^{\,\prime})\right] (n^\uparrow_{\vec{p}^{\,\prime}}-
n^\downarrow_{\vec{p}^{\,\prime}})-
 A(\vec{p},\vec{p}^{\,\prime})\right\}
\frac{ d^3 p^\prime}{(2 \pi)^3}- \nonumber \\
&&- \int\left\{ i\left[\frac{4 \pi a}{m}+
B(\vec{p}^{\,\prime},\vec{p})\right] (n^\uparrow_{\vec{p}}-
n^\downarrow_{\vec{p}})-
 A(\vec{p}^{\,\prime},\vec{p})\right\}
n_{\uparrow\downarrow}(t,\vec{r},\vec{p}^{\,\prime})
\frac{ d^3 p^\prime}{(2 \pi)^3} \,\,.
\label{eq:Diffkin}
\end{eqnarray}
Here $n^{\uparrow,\downarrow}_{\vec{p}} =
\left[\exp((\epsilon _p \mp H/2 -\epsilon _F)/T) +1\right]^{-1}$
are the equilibrium distribution
functions for up- and down-spin particles, respectively, $a$ is the s-wave
scattering length, and
$\epsilon _p =p^2 /2m$.
Throughout this paper we use energy units for the field
$H$ and we set $\hbar=1$. The two terms in the r.h.s. of equation
(\ref{eq:Diffkin}) correspond to direct and exchange two-particle
scattering
processes, described by the finite-temperature
many-body $T$-matrix,
\begin{eqnarray}
T(\vec{p}_1,\vec{p}_2;(g_0,\vec{g})) =
&-&\frac{4\pi a}{m} +\left(\frac{4\pi a}{m}\right)^2\int\left[
\frac{\pi i}{2} \left(1-n^\uparrow_{\vec{g}+\vec{k} }-n^\downarrow
_{\vec{g} - \vec{k}}\right)^2
\delta\left(g_0+\epsilon_F-\frac{g^2+k^2}{2m} \right) - \right. \nonumber \\
 &-&\left. {\cal P}
 \frac{1-n^\uparrow_{\vec{g}+\vec{k} }-n^\downarrow
_{\vec{g} - \vec{k}}     }
{2g_0 -g^{\,2}/m+ 2
\epsilon_F - k^2/m}
+{\cal P}\frac{m}{p^2_2-k^2}
\right] \frac {d^3k}{(2\pi)^3}\,\,\,,
\label{eq:Tequil}
\end{eqnarray}
where $(g_0, \vec{g})$ is the $4$-momentum of the center of mass, the
incoming particles have momenta $\vec{g}\pm \vec{p}_1$, and the
outgoing -- $\vec{g}\pm \vec{p}_2$.

In the kinetic equation (\ref{eq:Diffkin}), the functions
$A(\vec{p},\vec{p}^{\,\prime})$ and $B(\vec{p},\vec{p}^{\,\prime})$
contain all effects of second order in $ap_F$, $p_F =\sqrt{2m\epsilon _F}$.
In the case of
low temperature and small polarization, $\,\,T,H \ll \epsilon_F$, these
take the form,
\begin{equation}
A(\vec{p},\vec{p}^{\,\prime}) \approx \frac{ \pi a^2}{p}
\left(\frac{(p^2+{p^\prime}^2)}{m}- 4
\epsilon_F\right)\cdot\left(
n^\uparrow_{\vec{p}^{\,\prime}}+
n^\downarrow_{\vec{p}^{\,\prime}}-\frac{1}{1-{\rm exp} \left[
\frac{p^2+{p^\prime}^2}{2mT}-\frac{2\epsilon_F}{T}
\right]} \right)\,,
\label{eq:aaproxT}
\end{equation}
\begin{equation}
B(\vec{p},\vec{p}^{\,\prime}) \approx \frac{2a^2}{m} \left( -4p_F + |
\vec{p} - \vec{p}^{\,\prime}| \log \left|\frac{|\vec{p} -
\vec{p}^{\,\prime}| + |\vec{p} +
\vec{p}^{\,\prime}| }{|\vec{p} -
\vec{p}^{\,\prime}| - |\vec{p} +
\vec{p}^{\,\prime}|}\right|\right)\,.
\end{equation}
A more detailed derivation of the kinetic equations will be given in a longer
publication.

To compute the transverse spin-diffusion coefficient we will solve
the kinetic equations (\ref{eq:Diffkin}) and extract
the steady state transverse spin-current driven by
a constant magnetization gradient. For simplicity we will
consider the case in which
$\vec{M}_\bot (\vec{r}) \equiv \vec{M}_\bot (x)$ with
$\partial \vec{M}_\bot (x) /{\partial x} = const$.
We will also restrict ourselves to the
low temperature, weakly polarized limit,
$T, H \ll \epsilon _F$,
in which case a complete analytical solution is possible in the
s-wave approximation.

With these assumptions the solution of the kinetic equation (\ref{eq:Diffkin})
can be taken to be of the form, $n_{\uparrow \downarrow} (t, \vec{r} ,\vec{p})
= g(t,x,p) + f(t,x,p) \cos \psi$,
where $p=|\vec{p}|$ and $\cos \psi =\vec{p} \cdot \hat{\bf{x}} /p$.
Performing explicit integrations in equation (\ref{eq:Diffkin}) then leads to
the two coupled kinetic equations,
\begin{eqnarray}
\frac{\partial g}{\partial t} + \frac{p}{2m}\frac{\partial f}{\partial
x} + \frac {2a}{\pi m} i\int_0^\infty g(t,x,q) q^2 dq (\Nu{p} - \Nd{p})
&-& \frac{4 \pi a}{m}i(N_\uparrow- N_\downarrow)g = \nonumber \\
&=& -I_{rel}[g]-iI_{sr}[g]
\label{eq:legen0}
\end{eqnarray}
and
\begin{eqnarray}
\frac{\partial f}{\partial t} + \frac{p}{m} \frac{\partial g}{\partial
x} - \frac{4 \pi a}{m}i(N_\uparrow- N_\downarrow)f &=&
-\frac{I_{rel}[f\cos \psi]+iI_{sr}[f\cos\psi]}{\cos\psi} \equiv
\nonumber \\
&\equiv & - I_{rel1}[f]-iI_{sr1}[f]\,.
\label{eq:legen1}
\end{eqnarray}
In (\ref{eq:legen0}--\ref{eq:legen1}) $I_{rel}$ and $I_{sr}$
represent the relaxational and spin-rotation parts of the ``collision
integral'',
\begin{equation}
I_{rel}[W(\vec{p})]=W(\vec{p})\,  \int A(\vec{p},\vec{p}^{\,\prime})
\frac{ d^3 p^\prime}{(2 \pi)^3}- \int A(\vec{p}^{\,\prime},\vec{p})
W(\vec{p}^{\,\prime})
\frac{ d^3 p^\prime}{(2 \pi)^3}\,\,,
\label{eq:reltot}
\end{equation}
\begin{equation}
I_{sr}[W(\vec{p})]= -
W(\vec{p}) \int
B(\vec{p},\vec{p}^{\,\prime}) (n^\uparrow_{\vec{p}^{\,\prime}}-
n^\downarrow_{\vec{p}^{\,\prime}}) \frac {d^3 p^\prime}{(2 \pi)^3} +
 (n^\uparrow_{\vec{p}}-
n^\downarrow_{\vec{p}}) \int B(\vec{p},\vec{p}^{\,\prime})
W(\vec{p}^{\,\prime}) \frac {d^3 p^\prime}{(2
\pi)^3} \,\,.
\label{eq:spinrotcoll}
\end{equation}
As can be shown by analysing the eigenvalue spectrum of the relaxational
part of the collision integral $I_{rel1} [f]$ the
spin current decays to its steady state value in a microscopic time
scale, $\tau _\bot$, beyond which the time derivative terms in
equations (\ref{eq:legen0}--\ref{eq:legen1}) can be omitted. In addition, since
$\partial \vec{M} _\bot (x)/\partial x = const$ or equivalently
$\partial g /\partial x$ is independent of $x$, equation
(\ref{eq:legen1}) implies
$\partial f/\partial x = 0$. The solution of (\ref{eq:legen0})
can then be shown to be
\begin{equation}
g_0(x,p)=G(x) (\Nu{p}-\Nd{p}) \,.
\label{eq:zerosol}
\end{equation}
in which case the equation for $f(p)$ can be written as
\begin{equation}
v_F \frac{\partial G}{\partial x}(\Nu{p} - \Nd{p}) - i \Omega^{(1)}
f(p) = -I_{rel1}[f(p)]-iI_{sr1}[f(p)] .
\label{eq:difmicro}
\end{equation}
Here, $v_F$ is the Fermi velocity and
$\Omega^{(1)}=(8\pi a / m)M_\parallel \approx
2a p_F H/ \pi$ represents the leading correction to
the precession frequency of the spin current,
$J^-_\bot \equiv J^x_\bot - iJ^y_\bot = \int_0^\infty
f(p)p^3dp / 6\pi^2m $ ($M_\parallel$ is the longitudinal magnetization
induced by the field $H$).
Note that no such correction
appears in the equation for the transverse magnetization itself,
$M^-_\bot \equiv M^x_\bot -i M^y_\bot = \int_0^\infty
g(p) p^2 dp /4\pi ^2$, i.e., there is no renormalization of
Larmor precession, as expected from rotational invariance of the
inter-particle interactions.

To identify the diffusion coefficient we want to match the conventional form
\cite{Leggett,Helium3}
of the macroscopic constitutive relation,
\begin{equation}
J^-_\bot = - \frac{D_\bot}{1-i\xi} \frac{\partial}{\partial x}
M^-_\bot\,
\label{eq:difmacro}
\end{equation}
to the one implied by the microscopic equation (\ref{eq:difmicro}).
Here, $D_\bot$ is the transverse spin diffusion coefficient while
$\xi$ is referred to as the ``spin-rotation" parameter.
The non-zero imaginary part of the denominator in (\ref{eq:difmacro})
reflects the fact that, due to the spin-rotation effect, the spin current
is not parallel (in spin-space) to the driving magnetization gradient.
We recall that in the
relaxation-time approximation \cite{Leggett} $\xi \approx
\Omega ^{(1)} \tau _{relax}$, where $\tau _{relax}$ is the corresponding
relaxation time. We note from the outset that, except in the
high field regime, $H \gg T$, the relaxation-time approximation breaks down
and the above parametrization of $\xi$ is inapplicable.
We are now in position to compute $D_\bot$ and $\xi$ as functions of
$T$ and $H$. As we show below, the calculation is tractable
analytically in both ``high"- and ``low"-field limits:

\underline{High-Field Behavior, $H/T \gg (ap_F )(T/{\epsilon _F})$:}
In this case
the relaxational term in (\ref{eq:difmicro}) is small,
$I_{rel1}[f] \ll \Omega^{(1)}f$,
and the current $\vec{J}_\bot$ is almost perpendicular to the
magnetization gradient.
In this limit the solution of equation (\ref{eq:difmicro}), $f_D (p)$,
can be obtained iteratively:
\begin{equation}
f_D(p)=-i\frac{v_F}{\Omega^{(1)}}\frac{\partial G}{\partial x}
(\Nu{p}-\Nd{p}) - \frac{v_F}{(\Omega^{(1)})^2}\frac{\partial
G}{\partial x}I_{rel1}[\Nu{p}-\Nd{p}]-i\frac{v_F} {(\Omega^{(1)})^2}
\frac{\partial G}{\partial x} I_{sr1}[\Nu{p}-\Nd{p}]
\,\,.
\label{eq:highsol}
\end{equation}
Upon integration this leads to
\begin{equation}
J^-_\bot[f_D(p)]=-i\frac{v_F}{\Omega^{(1)}}\frac{\partial G}{\partial x}
J_\bot [\Nu{p}-\Nd{p}]\left\{ 1+ \frac{1}{\Omega^{(1)}}
\frac{J_\bot [I_{sr1}[\Nu{p}-\Nd{p}]]}{J_\bot [\Nu{p}-\Nd{p}]} -
 \frac{i}{\Omega^{(1)}}
\frac{J_\bot[I_{rel1}[\Nu{p}-\Nd{p}]]}{J_\bot [\Nu{p}-\Nd{p}]}\right\}
\,,
\label{eq:curhigh}
\end{equation}
where $J_\bot [W(p)] = \int W(p) p^3 dp/12\pi ^2 m$.
In (\ref{eq:highsol}) and (\ref{eq:curhigh}) we identify
$J_\bot [I_{sr1}[\Nu{p}-\Nd{p}]]/J_\bot [\Nu{p}-\Nd{p}] =
-\Omega ^{(2)}$ with the second order correction to the spin current
precession frequency, and
$J_\bot[I_{rel1}[\Nu{p}-\Nd{p}]]/J_\bot [\Nu{p}-\Nd{p}] =
\tau _D  (H,T)$ with the ``diffusion time".
We note that the ``diffusion time", $\tau _D $,
is not the relaxation time of the spin current usually involved in
the phenomenological discussions: the latter describes the relaxation of
the distribution function towards the steady state solution, $f_D (p)$ and
would arise as the eigenvalue of the relaxational part of the
collision operator (see below).
With these definitions,
(\ref{eq:difmacro}) and (\ref{eq:curhigh}) immediately yield,
\begin{eqnarray}
D_\bot&=&\frac{v_F^2 \tau_D}{3}= \frac{3 \pi v_F^2}{8 m a^2 ( H^2 +
4 \pi^2 T^2)}
\label{eq:diffhigh}\\
\xi&=& (\Omega^{(1)} + \Omega^{(2)}) \tau_D = \frac {9 H
v_F} {4 a (H^2 +
4 \pi^2 T^2)} [1 + \frac{2}{5 \pi}a p_F (1- \frac{4}{3} \log 2)].
\end{eqnarray}

The first thing to stress is that the
behaviour of $D_\bot$ in (\ref{eq:diffhigh}) differs significantly
from that of the longitudinal spin diffusion coefficient \cite{Miyake} ,
$D_\parallel \approx (v_F^2 / 8 \pi m a^2 T^2 ) C(-1/3)$,
where $C(-1/3) \approx 0.843$ is the Brooker-Sykes
coefficient~\cite{Brooker}.
The origin of this effect is
the different phase space restrictions associated with
the scattering processes leading to transverse and longitudinal spin
diffusion in the high-field limit.
While collisions responsible for longitudinal spin diffusion
are restricted to energies within $k_B T$ of each of the (``up" or ``down")
Fermi surfaces, those leading to transverse spin diffusion involve spin-flips
which can also take advantage of the full region enclosed {\em between}
the two surfaces. Thus, the expression for $D_\bot$ in (\ref{eq:diffhigh})
involves the phase-space for scattering generated both by the magnetic field
as well as by the thermal smearing of the individual Fermi surfaces.
As a result, the transverse
diffusion coefficient in (\ref{eq:diffhigh})
remains {\em finite} at $T=0$ as was predicted
by Meyerovich
\cite{Meyer85} and recently confirmed experimentally \cite{Candela}.

It should be noted that, even for $T \gg H$ (but still in the high field
limit, $H \gg (ap_F)T^2 /\epsilon _F$)
the transverse spin diffusion coefficient
$D_\bot \approx (3 v_F^2 / 32 \pi m a ^2 T^2 ) \approx 0.890
D_\parallel$
still differs from the longitudinal one. This is contrary to the
commonly implied belief~\cite{Mullin89,Karen3}
that the crossover to the isotropic regime of
spin diffusion occurs for $T\sim H$.
As we will see below $D_\bot$ and $D_\parallel$ become equal only
at much lower fields, $H\sim (ap_F) T^2 /\epsilon _F$.

\underline{Crossover and Low-Field Behavior, $H/T \stackrel{<}{\sim}
(ap_F ) T/\epsilon _F $:}
In this region,
$I_{rel1}[f(p)] \sim \Omega^{(1)}f(p) \gg I_{sr1}[f(p)]$
and, therefore, the spin-rotation terms in the collision integral may
be omitted. Furthermore, one may also set $H=0$ in evaluating
the functional $I_{rel1}$.
In terms of reduced variables, $\eta (p) = (\epsilon _p -\epsilon _F ) /T$
(and $f(\eta (p)) \equiv f(p)$),
$h=H/2T \ll 1$, and $\Gamma _0 = 2a^2m/3\pi$, the steady state
equation (\ref{eq:difmicro}) then becomes,
\begin{eqnarray}
-\frac{v_F}{T^2}\frac{\partial G}{\partial x}&&\left(\frac{1}{e^{\eta-h}+1}
-\frac{1}{e^{\eta+h}+1} \right) = 3\Gamma_0 f(\eta) \left( \eta^2 + \pi^2 -
\frac{i \Omega^{(1)}}{3 \Gamma_0 T^2} \right) - \nonumber \\
&&-2\Gamma_0
\int_{-\infty}^{\infty}(\eta+\sigma)f(\sigma)\left(\frac{1}
{e^\sigma+1}+
\frac{1}{e^{\eta+\sigma}-1} \right)d\sigma\,.
\label{eq:diffintermed}
\end{eqnarray}
Our analysis makes use of the
methods of Reference \cite{Brooker} and begins by
transforming (\ref{eq:diffintermed}) into the differential
equation,
\begin{equation}
F''(k) - \pi^2 \gamma^2 F(k) - \frac{2}{3} {\rm sech}^2 \pi k \,F(k) =
\frac{\pi h v_F}{3 \Gamma_0 T^2} \frac{\partial G}{\partial x}
\frac{\cos k h}{\cosh \pi k}
\label{eq:intermed2}
\end{equation}
for the function
$F(k) = \int_{-\infty}^{\infty}e^{i k \eta} f(\eta) \cosh (\eta /2) d\eta$.
Here
$\gamma^2=1- i \Omega^(1) /3 \pi^2 \Gamma_0 T^2 \approx
1 - 2i H \epsilon _F /\pi ^2 T^2 ap_F$.

The solutions to (\ref{eq:intermed2}) may be expressed in terms of
Gegenbauer polynomials~\cite{Gegen} with a complex index,
\begin{equation}
F(k) = \sum_{n=0}^\infty F_n \phi_n(k)\,\,\,\,\,,
\phi_n(k)=(\cosh \pi k)^{-\gamma}C^{\gamma +
\frac{1}{2}}_n(\tanh \pi k).
\label{eq:diffhomeigen}
\end{equation}
where
\begin{equation}
F_n = -\frac{v_F h}{\pi^2 \Gamma_0 T^2}\frac{\partial G}{\partial x}
\frac{g_n}{\frac{2}{3}+ (\gamma+n)(\gamma+n+1)}\,.
\label{eq:inhomeigen}
\end{equation}
and
\begin{equation}
g_n=\frac{\pi^{5/2}(n+\gamma + \frac{1}{2})\Gamma(\gamma+\frac{1}{2})}
{3 \cos \frac{\pi \gamma}{2}\, \Gamma(\frac{1}{2}-\frac{n}{2})
\Gamma(1+\gamma+\frac{n}{2})\Gamma(1+\frac{\gamma}{2}+\frac{n}{2})
\Gamma(\frac{1}{2}-\frac{\gamma}{2}-\frac{n}{2})}.
\end{equation}
Note, that since the functions $f(\eta)$ and $F(k)$ are even and
$\phi_n(-k) = (-1)^n \phi_n(k)$, only even $n$ terms contribute to the
sum in (\ref{eq:diffhomeigen}).

The transverse spin current can then be evaluated as
\begin{equation}
J_\bot^-=(T
p_F^2/ 12\pi^2) \int_{-\infty}^\infty dk F(k)/\cosh {\pi k} = (T p_F^2
/ 12\pi^2) \sum_{n=0}^\infty F_n a_n\,,
\end{equation}
with
\begin{equation}
a_n =\frac{\pi \Gamma(\gamma+1) \Gamma(2\gamma+n+1)}{n{\rm !} \cos\frac{\pi
\gamma}{2}\, \Gamma(2\gamma+1) \Gamma(\frac{1}{2}-\frac{n}{2})
\Gamma(\gamma+\frac{n}{2}+1)\Gamma(\frac{\gamma}{2}+\frac{n}{2}+1)
\Gamma(\frac{1}{2}-\frac{n}{2}-\frac{\gamma}{2})}\,.
\end{equation}
This allows us to write the final expression,
\begin{equation}
\frac{D_\bot}{1-i\xi} =
\frac{1}{4 \pi}\frac{v_F^2}{m a^2 T^2} \sum_{n=0}^\infty
\frac{g_n a_n}{\frac{2}{3} +(\gamma+n)(\gamma+n+1)}\,,
\label{eq:intermedanswer}
\end{equation}
from which we can extract $D_\bot$ and the
spin-rotation parameter $\xi$.
The low field limit, $H<<(ap_F) T^2 /\epsilon _F$ corresponds to
$\gamma \rightarrow 1$ in which case (\ref{eq:intermedanswer}) yields,
$D_\bot = D_{\parallel} =
(v_F^2/8 \pi m a^2 T^2) C(-1/3)$.
In this limit the spin-rotation
parameter is given by
$\xi \approx 0.139 H\epsilon _F/T^2 (ap_F) << 1$, reflecting the fact that
the spin current becomes parallel (in spin space)
to the driving magnetization gradient.
The detailed behavior of $D_\bot$ as a function of $T/H$,
which clearly displays the two crossovers discussed above,
is shown in Figure 1.

Although the calculations presented above are restricted to
low densities,
we expect that our qualitative arguments, based on the existence of {\em two}
independent parameters, $\xi$ and $H/T$, should also hold
in the Fermi liquid regime, $T,H << \epsilon _F$.
The $s$-wave approximation,
which made our analytical calculations possible, raises more serious
issues, especially concerning the detailed behavior of $D_\bot$ in the
crossover regions. In particular, we would expect that the region in
which $D_\bot \propto T^{-2}$ with $D_\bot \neq D_\parallel$ should shrink.

Some comments are in order concerning the possible
relevance of our findings to experiment.
First we note that the available experimental data
in weakly polarized $^3{\rm He}$ ~\cite{Candela}
deviate systematically from the naive theoretical fit
which uses a single adjustable parameter ($T_a$ in \cite{Candela})
to cover the entire temperature range including both, $\xi >1$ and $\xi < 1$.
Much better agreement is obtained by restricting the fit to
the $\xi \leq 1$ region with an overall prefactor smaller than
the one implied by fitting to the value of $D_\parallel$ in the
low field, high temperature regime (we use for the inverse overall
prefactor the value $A \approx 7.7 \times 10^5 \rm{sec}/\rm{cm}^2 \rm{K}^2$
instead of $5.8 \times 10^5 \rm{sec}/\rm{cm}^2 \rm{K}^2$ as in
Reference \cite{Candela} and for the ``anisotropy temperature'' $ T_a \approx
12.5$ mK instead of 16.4 mK). This is consistent with
our picture, with $D_\bot < D_\parallel$ for
$\xi \geq 1$ and $T >> H$. In addition,
although it appears that the region between the
two crossovers cannot be clearly identified --  most likely due to large
Fermi liquid renormalization effects -- the isotropic limit is indeed reached
in the regime $\xi < 1$~\cite{Candela}.
In principle,
our calculations should be more relevant to the measurements in
dilute $^3{\rm He}$ --$^4{\rm He}$ mixtures.
Although in the available data
(for $.18 \% \,\,^3{\rm He}$)
the crossover to the
isotropic limit occurs for $\xi \sim 1$ with $H << T$, the temperature is
not sufficiently far below $\epsilon _F$ and, moreover, the polarization is
somewhat high, $\sim 25 \%$.
Nevertheless,
for $\xi > 1$ $D_\bot \propto D_\parallel$ with
the ratio $D_\bot / D_\parallel$ slightly less than
unity~\cite{Candela91,Nunes}.
Also, the measured $T$ dependence of the ``spin-rotation"
parameter, $\xi$, near the crossover to the isotropic (``low-field")
limit
is qualitatively consistent with our results in
both the data of
reference \cite{Candela91} and those obtained in the degenerate
regime of more concentrated solutions ($2.6 \%\,\, ^3{\rm
He}$)~\cite{Owers} with lower
polarization ($\sim 2 \%$). In both situations, the crossover to the
isotropic regime can be clearly distinguished.
However, in the former case the
polarization was rather high ($\sim 25 \%$) and the ``low-field'' crossover
itself occurs beyond the degenerate limit.
Also, there is a large discrepancy in the magnitude of
the shift of $\xi T^2$ in the latter case (see Figure 2)
which can be attributed to Fermi liquid renormalizations anticipated
in high concentration solutions.
To sharpen the identification of two crossovers
the data of reference \cite{Owers} should be extended to lower temperatures
(to study the $H/T \sim 1$ behavior). Quantitative comparisons could be made
only in the more dilute case of Reference \cite{Candela91} where lower
field and lower temperature experiments should be performed.

We close with a brief comment about the time scale for the relaxation
to the steady state solution.
This relaxation proceeds exponentially with
a transverse relaxation time $\tau_\bot$ defined (in the appropriate
reference frame)
by
\begin{equation}
-\frac{1}{\tau_\bot}f(p) + i \tilde{\Omega}^{(2)}f(p) =
-I_{rel1}[f]-iI_{sr1}[f]\,,\,\,\,\,\,\,
\tilde{\Omega}^{(2)} \sim \Omega^{(2)}\,.
\label{eq:transient}
\end{equation}

We have only solved this equation in the
limit $H \gg T^2/\epsilon_F$ in which case the spin-rotation term
$I_{sr1}$ results in the rapid oscillations of the distribution
function $f(p)$ with a characteristic frequency
$\sim \Omega ^{(2)} >> 1/\tau _\bot$.
Upon averaging these oscillations over time scales short compared to
$\tau _\bot$, (\ref{eq:transient}) reduces to an
eigenvalue problem for the collision operator, $I_{rel1}$,
$f(p)/{\tau_\bot}=I_{rel1}[f(p)]$. The result of this calculation, plotted
in Figure 3, illustrates two important points: first of all,
$\tau _\bot$ is different from the ``diffusion time", $\tau _D$. Moreover,
its $T,H$ dependence cannot be reduced to the ``conventional" form,
$\tau _\bot ^{-1} \propto A T^2 + B H^2$.

\acknowledgements

We are grateful to V. A. Brazhnikov, A. E. Meyerovich, and K. A. Musaelian
for helpful and enjoyable discussions and to D. Candela for providing
us with the data of References \cite{Candela} and \cite{Candela91}
in a convenient form.
This work was supported in part by ONR Grant \# N00014-92-J-1378.

\figure{Spin diffusion coefficient in the high-field region
$H \stackrel{>}{\sim} T$. The transverse
spin diffusion coefficient $D_\bot$ (equation (\ref{eq:diffhigh})),
the longitudinal coefficient $D_\parallel$,
and the naive fit  $\left(1/D_\bot^{(0)}+1/D_\parallel\right)^{-1}$
for $D_\bot$ ($D_\bot^{(0)}=3\pi v_F^2/8ma^2H^2$ being the limiting
value of $D_\bot$ at $T \rightarrow 0$) are represented by
the solid, dashed and dotted lines, respectively.
The inset shows the ratio $D_\bot/D_\parallel$
near the second (low-field) crossover, $T\sim T_c=\sqrt{H\epsilon_F/a p_F}$.
\label{fig:diffhigh}}

\figure{The low-field behaviour of the spin-rotation parameter $\xi$
in the concentrated $^3{\rm He}$--$^4{\rm He}$ mixture. The points
represent the experimental data of Reference~\cite{Owers}, the dashed
line -- the $T^{-2}$ fit for the behaviour of $\xi$ below the crossover,
and the solid line is our theoretical result for the $s$-wave
approximation (with the values of $T$ and $\xi$ scaled by appropriate
factors). Note, that the latter is {\em not} a straight line. On
the inset, the theoretical curve for $\xi$ at  $T\sim
T_c=\sqrt{H\epsilon_F/a p_F}$ is plotted in a different scale.
\label{fig:spinrot}}

\figure{The relaxation time of the transverse spin current plotted as
a function of $T/ H$. The solid line corresponds to the exact
eigenvalue of collision integral and the dotted
line -- to the fit $\tau_\bot^{-1}= AT^2+BH^2$, which differs from the
exact value
in the crossover region $ H \sim T$. The dashed line represents the
diffusion relaxation time $\tau_D$ (see equation (\ref{eq:diffhigh})).
\label{fig:tauexact}}


\begin{references}
\bibitem{Gegen} The functions $\phi _n (k)$ are solutions to the homogeneous
equation $F''(k) -[\pi ^2 \gamma ^2 + (\gamma +n)(\gamma +n +1)
sech ^2 \pi k ]F(k) =0$.
\end{references}
\end{document}